\documentclass[aps,preprint]{revtex4}
\usepackage{amsfonts}
\usepackage{amsmath}
\usepackage{amssymb}
\usepackage{graphicx}

\input{tcilatex}

\begin{document}

\title{Steady entanglements in bosonic dissipative networks}
\author{G. D. de Moraes Neto$^{1}$, W. Rosado$^{1}$, F. O. Prado$^{2}$, and
M. H. Y. Moussa$^{1}$}
\affiliation{$^{1}$Instituto de F\'{\i}sica de S\~{a}o Carlos, Universidade de S\~{a}o
Paulo, Caixa Postal 369, 13560-970, S\~{a}o Carlos, S\~{a}o Paulo, Brazil}
\affiliation{$^{2}$Universidade Federal do ABC, Rua Santa Ad\'{e}lia 166, Santo Andr\'{e}%
, S\~{a}o Paulo 09210-170, Brazil}

\begin{abstract}
In this letter we propose a scheme for the preparation of steady
entanglements in bosonic dissipative networks. We describe its
implementation in a system of coupled cavities interacting with an
engineered reservoir built up of three-level atoms. Emblematic bipartite ($%
Bell$ and $NOON$) and multipartite ($W$-class) states can be produced with
high fidelity and purity.
\end{abstract}

\pacs{PACS numbers: 32.80.-t, 42.50.Ct, 42.50.Dv}
\maketitle

The development of strategies to prepare nonclassical states of the
radiation and the vibrational fields \cite{PNS} and, in particular, to
protect them against decoherence ---by means of decoherence-free subspaces 
\cite{DFS}, dynamical decoupling \cite{DD}, and reservoir engineering \cite%
{PCZ,RE}--- has long played a significant role in quantum optics. On the
conceptual side, the need for these states stems from their use in the study
of fundamental quantum processes, for instance to track decoherence \cite%
{Decoherence} and the quantum-to-classical transition \cite{QC}. At the
practical level, mastery in handling these states is sought in the rapidly
growing field of quantum information theory ---which has recently mobilized
practically all areas of low-energy physics--- so as to implement the logic
operations required for quantum computation and communication \cite{Livro}.

The proposition of schemes that enable the generation of nonclassical
equilibrium states has been one of the main tasks in the field of quantum
information science. In this regard, the reservoir engineering technique
proposed in 1996 \cite{PCZ} and experimentally demonstrated four years
latter in a trapped-ion system \cite{IT} signalled an important step towards
the implementation of quantum information processes \cite{Livro}. Moreover,
the protection of a particular state demands the (not always easy)
engineering of a specific interaction which the system of interest is forced
to perform with other, auxiliary, quantum systems. In the most important
case of preparing and protecting entangled states, recent theoretical
protocols, also based on engineering the decay process \cite{re}, have all
been shown possible with only dissipation as a resource. However, most of
these theoretical schemes concentrate on the preparation of atomic maximally
entangled states of two qubits \cite{2atom}, the $W$ state of three qubits 
\cite{3atom} or atomic multipartite entangled states \cite{Natom}.
Furthermore, we could mention schemes for preparing non-maximal steady
entanglements of two or three oscillators coupled to a common reservoir \cite%
{2harmonic}, while, more recently, a proposal has been advanced to engineer
a common squeezed reservoir for an ensemble of oscillators that has genuine
multipartite entanglement \cite{Nharmonic}.

In this Letter, based on our previous work \cite{FockEPL} concerning a
scheme to obtain Fock equilibrium states in a single cavity mode, we present
a strategy to produce high-fidelity steady entanglements in coupled quantum
harmonic oscillators (QHOs). Our protocol can be readily understood in terms
of the map between the natural oscillators and the normal-mode basis, in
which a prepared steady Fock state in a giving normal-mode oscillator must
correspond to a steady entangled state in the natural basis. The engineered
reservoir is built from a selective Jaynes-Cummings interaction \cite%
{WilsonJOPB}, in accordance with the prescription in \cite{FockEPL},
prompting the emergence of an engineered selective Liouvillian to govern the
normal-mode dynamics, alongside the Liouvillian accounting for the natural
loss mechanisms. We stress, from a practical perspective, that atomic
reservoirs have for some time been used for the preparation of the cavity
vacuum state \cite{RMP}. Moreover, this has been theoretically explored, in
close relation to the reservoir engineering technique \cite{PCZ}, for the
generation of an Einstein-Podolsky-Rosen steady state comprising two
squeezed modes of a high-finesse cavity \cite{BR}. We note that, the \textit{%
atomic reservoir} can be implemented in other contexts of atom-field
interaction, such as trapped ions \cite{Rafael} and circuit QED \cite{Nori},
where the required beam of atoms is simulated by a pulsed classical field.
In trapped ions, the classical field is used intermittently to couple the
vibrational mode with internal electronic states, while in circuit QED, it
is used to bring a Cooper-pair box into resonance with the mode of a
superconducting strip.

In our protocol, the steady state is driven by a sum of three engineered
Lindbladians, two of which act upon selected subspaces of the normal mode
space, one for photon emission and the other for photon absorption, within
the corresponding selected subspaces. The third Lindbladian is associated
with (non-selective) photon absorption by the normal mode, to counterbalance
the inevitable emission to the natural (nonengineered) environment. The
selective Lindbladians are built up from engineered selective
Jaynes-Cummings (JC) Hamiltonians, while the nonselective Lindbladian
follows from the usual JC interaction. We present here a brief review of the
steps in the derivation of the master equation of a bosonic network, as
developed in \cite{Temp}. We note that a given topology of a network
composed of $N$ QHOs is defined by the way the oscillators are coupled
together, the set of coupling strengths $\{\lambda _{mn}\}$ and their
natural frequencies $\{\omega _{m}\}$. Here we assume the general scenario,
where each oscillator is coupled to its own reservoir, instead of the
particular situation where the whole network is coupled to a common
reservoir. From here on, setting the indices $m$ and $n$ to run from $1$ to $%
N$, the Hamiltonian $H=H_{S}+H_{R}+H_{I}$ modelling this network accounts
for the $N$ coupled oscillators, given by ($\hbar =1$):%
\begin{equation}
H_{s}=\sum_{m}\left[ \omega _{m}a_{m}^{\dagger }a_{m}+\sum_{n(\neq
m)}\lambda _{mn}(a_{m}^{\dagger }a_{n}+a_{m}a_{n}^{\dagger })\right] ,
\label{Hs}
\end{equation}%
where the $N$ distinct reservoirs, $H_{R}=\sum_{m}\sum_{k}\omega
_{mk}b_{mk}^{\dagger }k_{mk}$, are each composed of an infinite set of $%
\{k\} $ modes, and the coupling between the QHOs and their respective
reservoirs: $H_{I}=\sum_{m}\sum_{k}V_{mk}(b_{mk}^{\dagger
}a_{m}+b_{mk}a_{m}^{\dagger })$. In the above, $a_{m}^{\dagger }$($a_{m}$)
is the creation (annihilation) operator associated with the $mth$ network
oscillator ($\omega _{m}$), which is coupled to the $nth$ oscillator with
strength $\lambda _{mn}$ and to the $mth$ reservoir with strength $V_{mk}$.
The $kth$ reservoir mode $\omega _{mk} $ is described by the creation
(annihilation) operator $b_{mk}^{\dagger }$($b_{mk}$). To derive the master
equation from Hamiltonian $H$, we first rewrite $H_{S}$ in a matrix form, $%
H_{s}=\sum_{m,n}a_{m}^{\dagger }\mathcal{H}_{mn}a_{m}$, the elements being
given by $\mathcal{H}_{mn}=\omega _{m}\delta _{mn}+\lambda _{mn}(1-\delta
_{mn}).$ The diagonalization of $\mathcal{H}$ is thus performed through the
canonical transformation $A_{m}=\sum_{n}C_{mn}a_{n}$, where the coefficients
of the $mth$ line of matrix $C$ define the eigenvectors associated with the
eigenvalues $\overline{\omega }_{m}$ of $\mathcal{H}$. $C$ being an
orthogonal matrix, $C^{T}$ $=C^{-1}$, the commutation relations $\left[
A_{m,}A_{n}^{\dagger }\right] =\delta _{mn}$ and $\left[ A_{m,}A_{n}\right]
=0$ follow, enabling the Hamiltonian $H$ to be rewritten as $\ H=H_{0}+V$,
where $H_{0}=\sum_{m}\left[ \overline{\omega }_{m}A_{m}^{\dagger
}A_{m}+\sum_{k}\omega _{mk}b_{mk}^{\dagger }k_{mk}\right] $ and $%
V=\sum_{m,n}\sum_{k}V_{mk}(b_{mk}^{\dagger }A_{n}+b_{mk}A_{n}^{\dagger })$.
With the diagonalized Hamiltonian $H_{0}$, we are ready to introduce the
interaction picture, defined by the transformation $U(t)=\exp (-\imath
H_{0}t),$ in which $V_{I}(t)=\sum_{m,n}(\mathcal{O}_{mn}(t)A_{n}^{\dagger }+%
\mathcal{O}_{mn}^{\dagger }A_{n}),$ with the bath operator $\mathcal{O}%
_{mn}(t)=C_{nm}\sum_{k}V_{mk}\exp [-\imath (\omega _{mk}-\overline{\omega }%
_{m})t]b_{mk}.$ Assuming the interactions between the oscillators and the
reservoirs to be weak enough, we perform a second-order perturbation
approximation, followed by tracing out the reservoir degrees of freedom. We
also assume a Markovian reservoir, where the time-dependent density operator
of the network can be factorized from the reservoir: $\rho _{1\ldots
N}(t)\otimes \rho _{R}(0).$

Next, we assume that the reservoir frequencies are sufficiently closely
spaced to allow a continuum summation and, as usual, that the coupling
strength $V_{m}(\overline{\omega })$ and the density of states $\sigma (%
\overline{\omega })$ of the $mth$ reservoir are slowly varying functions.
Moreover, assuming Markovian white noise reservoirs, where the damping rates
read $\gamma _{m}(\overline{\omega }_{k})=\gamma _{m\text{ }}$, the average
excitation of the reservoir associated with the $mth$ oscillator$\ $is $%
\left\langle b_{m}^{\dagger }(\overline{\omega }_{k})b_{m}(\overline{\omega }%
_{k})\right\rangle =\overline{n}_{m}(\overline{\omega }_{k})=\overline{n}%
_{m} $ and the cross-decay terms $\gamma _{mn\text{ }}$are null \cite{Temp},
we obtain the normal mode master equation 
\begin{equation}
\frac{d\rho }{dt}=-\imath \lbrack \overline{H},\rho ]+\mathcal{L}\rho ,
\label{eq1}
\end{equation}%
where $\overline{H}=\sum_{m}\overline{\omega }_{m}A_{m}^{\dagger }A_{m}$ and
the Liouvillian 
\begin{align}
\mathcal{L}\rho & =\dsum_{m}\frac{\gamma _{m}}{2}(1+\overline{n}_{m})\left(
2A_{m}\rho A_{m}^{\dagger }-\rho A_{m}^{\dagger }A_{m}-A_{m}^{\dagger
}A_{m}\rho \right) \\
& +\dsum_{m}\frac{\gamma _{m}}{2}\overline{n}_{m}\left( 2A_{m}^{\dagger
}\rho A_{m}-\rho A_{m}A_{m}^{\dagger }-A_{m}A_{m}^{\dagger }\rho \right) .
\end{align}%
Our strategy to produce equilibrium entanglements in a bosonic dissipative
network demands an \textit{engineered selective} Liouvillian, to be added to
the master equation (\ref{eq1}), having the following structure:%
\begin{align}
\mathcal{L}_{eng}\rho & =\dsum_{m}\frac{\Gamma _{m\ell }}{2}\left( 2A_{m\ell
}\rho A_{m\ell }^{\dagger }-\rho A_{m\ell }^{\dagger }A_{m\ell }-A_{m\ell
}^{\dagger }A_{m\ell }\rho \right)  \notag \\
& {\small +}\dsum_{m}\frac{\Gamma _{m\ell ^{\prime }}}{2}\left( 2A_{m\ell
^{\prime }}^{\dagger }\rho A_{m\ell ^{\prime }}-\rho A_{m\ell ^{\prime
}}A_{m\ell ^{\prime }}^{\dagger }-A_{m\ell ^{\prime }}A_{m\ell ^{\prime
}}^{\dagger }\rho \right)  \notag \\
& {\small +}\dsum_{m}\frac{\Gamma _{m}}{2}\left( 2A_{m}\rho
A_{A_{m}^{\dagger }}-\rho A_{m}^{\dagger }A_{m}-A_{m}^{\dagger }A_{m}\rho
\right) ,  \label{sele}
\end{align}%
where $A_{m\ell }^{\dagger }=\left\vert \ell +1\right\rangle \left\langle
\ell \right\vert $ $(A_{m\ell }=\left\vert \ell \right\rangle \left\langle
\ell +1\right\vert )$ is a selective creation (annihilation) operator acting
on the Fock subspace $\left\{ \left\vert \ell \right\rangle ,\left\vert \ell
+1\right\rangle \right\} $ of the $mth$ normal mode. The engineered
Liouvillians associated with the effective decay rates $\Gamma _{m\ell }$
and $\Gamma _{m\ell ^{\prime }}$ account for selective emission and
absorption terms, while the engineered Liouvillian associated with $\Gamma
_{m}$ represents an additional cooling term. This latter Liouvillian must be
taken into account, as will become clear later, only when preparing a
required equilibrium state where at least one of the normal modes, the $n$%
th, is in the vacuum state. If the normal modes are not degenerate and the
effective decay rates satisfy $\Gamma _{m\ell },\Gamma _{m\ell ^{\prime
}},\Gamma _{n\neq m}\gg \gamma _{m}$, with the additional condition $\ell
=\ell ^{\prime }+1$ (needed to generate a Fock state in a given normal
mode), the full master equation $\dot{\rho}=-\imath \lbrack \overline{H}%
,\rho ]+\mathcal{L}\rho +\mathcal{L}_{eng}\rho $ leads to a steady Fock
state $\left\vert \ell \right\rangle .$ The idea is to search for steady
Fock states in the normal mode basis that correspond to a steady
entanglement when mapped back to the natural oscillator basis. We will
explore this later on, when cases of symmetric networks will be studied. To
illustrate this protocol, we first address two significant cases: the $Bell$
and $NOON$ states in two-coupled-cavity system. We show that these states
are obtained by the generation of a single-excitation Fock state in a given
normal mode, and this requires only one selective Lindbladian instead of the
three engineered Lindbladians in Eq. (\ref{sele}). We emphasize that the use
of all three terms in Eq.(\ref{sele}) improves the fidelity of the target
steady state, but they are only essential to reach more excited entangled
states (this will be clarified in the simulations ahead).

It is straightforward to verify the mapping between the normal basis
(labeled $M$) and the natural oscillator basis (labeled $m$), which in the
cases of interest is reduced to the $Bell$ states $\left\vert
1,0\right\rangle _{M}=\frac{1}{\sqrt{2}}(\left\vert 1,0\right\rangle
_{m}+\left\vert 0,1\right\rangle _{m})$ and\ $\left\vert 0,1\right\rangle
_{M}=\frac{1}{\sqrt{2}}(\left\vert 1,0\right\rangle _{m}-\left\vert
0,1\right\rangle _{m})$, and the $NOON$ state $\left\vert 1,1\right\rangle
_{M}=\frac{1}{\sqrt{2}}(\left\vert 2,0\right\rangle _{m}-\left\vert
0,2\right\rangle _{m})$. Considering two nonideal coupled cavities (labeled $%
i,j=1,2)$ with degenerate frequencies $\omega $, coupling strength $\lambda $%
, decay rates $\gamma $, and the same temperature $T=\hbar \omega /k_{B}\ln %
\left[ \left( 1+\bar{n}\right) /\bar{n}\right] $, described by the
Hamiltonian $H_{c}=\omega \sum_{i}a_{i}^{\dagger }a_{i}+\lambda \sum_{i\neq
j}a_{i}^{\dagger }a_{j}$, which is diagonalized through the operators $%
A_{\pm }^{\dagger }=(a_{1}^{\dagger }\pm a_{2}^{\dagger })/\sqrt{2}$, we get
the particular master equation (\ref{eq1}): 
\begin{align}
d\rho /dt& =-\imath \left[ \overline{H},\rho \right]  \notag \\
& +(\gamma /2)(1+\overline{n})\sum_{\alpha =\pm }\left( 2A_{\alpha }\rho
A_{\alpha }^{\dagger }-\rho A_{\alpha }^{\dagger }A_{\alpha }-A_{\alpha
}^{\dagger }A_{\alpha }\rho \right)  \notag \\
& +(\gamma /2)\overline{n}\sum_{\alpha =\pm }\left( 2A_{\alpha }^{\dagger
}\rho A_{\alpha }-\rho A_{\alpha }A_{\alpha }^{\dagger }-A_{\alpha
}A_{\alpha }^{\dagger }\rho \right) ,
\end{align}%
where $\overline{H}=\sum_{\alpha =\pm }\overline{\omega }_{\alpha }A_{\alpha
}^{\dagger }A_{\alpha }$ and $\overline{\omega }_{\pm }=\omega \pm \lambda $.

The required selective Lindbladian can be constructed by following the
protocol presented in Ref. \cite{FockEPL}, extended to obtain selective
interactions in the Fock space of the normal modes. To this end, we consider
a beam of three-level atoms going through only one of the cavities (for
example, $i=1$), helped by two laser beams, $\omega _{1}$ and $\omega _{2}$,
to interact with the normal mode through the Hamiltonian $H=\frac{\Omega _{0}%
}{\sqrt{2}}\sigma _{ig}(A_{-}\limfunc{e}\nolimits^{-i\Delta _{-}t}+A_{+}%
\limfunc{e}\nolimits^{-i\Delta _{+}t})+\Omega _{1}\sigma _{ig}\limfunc{e}%
\nolimits^{i\Delta _{1}t}+\Omega _{2}\sigma _{ie}\limfunc{e}%
\nolimits^{-i\Delta _{2}t}+H.c.,$ where $\sigma _{rs}=\left\vert
r\right\rangle \left\langle s\right\vert $, $r$ and $s$ labelling the atomic
states involved, and $\Delta _{\pm }=\overline{\omega }_{\pm }-\omega _{ig}$%
, $\Delta _{1}=\omega _{ig}-\omega _{1}$, and $\Delta _{2}=\omega
_{2}-\omega _{ie}$, with $\omega _{i\ell }=\omega _{i}-\omega _{\ell }$ ($%
\ell =g,e$). For $\lambda \gg \Omega _{0},$ we have a strongly off-resonant
regime and, under the RWA, it follows that only one of the normal modes
effectively interacts with the atom \cite{Roversi}. We choose for example
the mode $\overline{\omega }_{+}$ to be almost resonant with the $%
g\longleftrightarrow i$ transition and henceforth we will omit the index of
the normal mode, such that $A_{+}=A$ and $\Delta _{+}=\Delta $. It is
straightforward to verify that the conditions $\Omega _{0}\sqrt{n+1}\ll
\Delta $ and $\Omega _{j}\ll $ $\Delta _{j}$ ($j=1,2$) lead to the effective
interaction (\cite{James}) $H_{eff}=\left( \xi A^{\dag }A-\varpi _{g}\right)
\sigma _{gg}+\varpi _{e}\sigma _{ee}+\left( \zeta A^{\dagger }\limfunc{e}%
\nolimits^{i\delta t}\sigma _{ge}+H.c.\right) ,$ where $\varpi
_{g}=\left\vert \Omega _{1}\right\vert ^{2}/\Delta _{1}$ and $\varpi
_{e}=\left\vert \Omega _{2}\right\vert ^{2}/\Delta _{2}$ stand for frequency
level shifts due to the action of the classical fields, whereas the
strengths $\xi =\left\vert \Omega _{0}\right\vert ^{2}/\Delta \sqrt{2}$ and $%
\zeta =\sqrt{2}\Omega _{0}^{\ast }\Omega _{2}\left( \Delta ^{-1}+\Delta
_{2}^{-1}\right) /4$ stand respectively for off- and on-resonant atom-field
couplings to be used to engineer the required selective interactions.
Finally,\textbf{\ }$\delta =\Delta -\Delta _{2}$\textbf{\ }refers to a
convenient detuning to be used to get selectivity.

We next perform the unitary transformation $U=\exp\left\{ -i\left[ \left(
\xi A^{\dag}A+\varpi_{g}\right) \sigma_{gg}+\varpi_{e}\sigma_{ee}\right]
t\right\} $, which takes $H_{eff}$ into the form $V_{eff}=\tsum
\nolimits_{n=1}^{\infty}\zeta_{n}\left\vert n+1\right\rangle \left\langle
n\right\vert \sigma _{ge}\limfunc{e}\nolimits^{i\phi_{n}t}+H.c.$, with $%
\zeta_{n}=\sqrt {n+1}\zeta$ and $\phi_{n}=\left( n+1\right)
\xi+\delta-\varpi_{g}-\varpi _{e}$. Thus, under the strongly off-resonant
regime, $\xi\gg\sqrt {k+2}\left\vert \zeta\right\vert $, and the condition $%
\phi_{\ell}=0$, which is easily satisfied by imposing $\left( m+1\right) \xi$
$=$ $\varpi_{g}$ $\gg\delta=\varpi_{e}$, such that $\left\vert
\Omega_{1}\right\vert =\sqrt{\left( m+1\right) \Delta_{1}/\Delta}\left\vert
\Omega_{0}\right\vert \gg\sqrt{\Delta_{1}/\Delta_{2}}\left\vert
\Omega_{2}\right\vert $, we readily eliminate, via RWA, all the terms
proportional to $\zeta_{n}=\sqrt{n+1}\zeta$ summed in $V_{eff}$, except $%
n=\ell$, bringing about the selective interaction $\mathcal{H}_{1}=\left(
\zeta_{\ell}\left\vert \ell+1\right\rangle \left\langle \ell\right\vert
\sigma_{ge}+H.c.\right) ,$ which produces the desired selective $%
g\leftrightarrow e$ transition within the Fock subspace $\left\{ \left\vert
\ell\right\rangle ,\left\vert \ell+1\right\rangle \right\} $. The excellent
agreement between this effective selective interaction and the full
Hamiltonian has been analyzed in detail in Ref. \cite{WilsonJOPB}.

Next, following the reasoning in Refs. \cite{BR,FockEPL} for atomic
reservoir engineering, and assuming all the atoms prepared in the excited
state $\left\vert e\right\rangle $, with the laser detuning $\Delta _{j}$
adjusted to produce $\ell ^{\prime }=0$ (i.e. $\phi _{\ell ^{\prime }}=0$),
we obtain the master equation 
\begin{align}
\frac{d\rho }{dt}& {\small =}\frac{\Gamma _{0}}{2}\left( 2A_{0}^{\dag }\rho
A_{0}-\rho A_{0}A_{0}^{\dag }-A_{0}A_{0}^{\dag }\rho \right)   \notag \\
& +(\gamma /2)(1+\overline{n})\sum_{\alpha =\pm }\left( 2A_{\alpha }\rho
A_{\alpha }^{\dagger }-\rho A_{\alpha }^{\dagger }A_{\alpha }-A_{\alpha
}^{\dagger }A_{\alpha }\rho \right)   \notag \\
& +(\gamma /2)\overline{n}\sum_{\alpha =\pm }\left( 2A_{\alpha }^{\dagger
}\rho A_{\alpha }-\rho A_{\alpha }A_{\alpha }^{\dagger }-A_{\alpha
}A_{\alpha }^{\dagger }\rho \right) ,  \label{7}
\end{align}%
where only the selective absorptive term in (\ref{sele}) is engineered, with
the effective rate $\Gamma _{0}=r\left( \zeta _{0}\tau \right) ^{2}$, $r$
being the atomic arrival rate and $\tau $ the average time during which each
atom crosses the cavity. The other two Lindbladians in Eq. (\ref{sele}), the
selective emission and the cooling terms, can be built as follows: by
preparing the atoms in the ground state $\left\vert g\right\rangle $ and
tuning $\Delta _{j}$ to produce $\ell =1$, we obtain \ the selective
Lindbladian $\frac{\Gamma _{1}}{2}\left( 2A_{1}\rho A_{1}^{\dagger }-\rho
A_{1}^{\dagger }A_{1}-A_{1}^{\dagger }A_{1}\rho \right) $ and, by switching
off the laser field and using a beam of two-level atoms prepared in the
ground state, resonant with one of the the normal modes $\overline{\omega }%
_{\pm }$, we get the cooling term $\frac{\Gamma _{\pm }}{2}\left( 2A_{\pm
}\rho A_{\pm }^{\dagger }-\rho A_{\pm }^{\dagger }A_{\pm }-A_{\pm }^{\dagger
}A_{\pm }\rho \right) .$

\begin{figure}
\includegraphics[width=1.0\columnwidth]{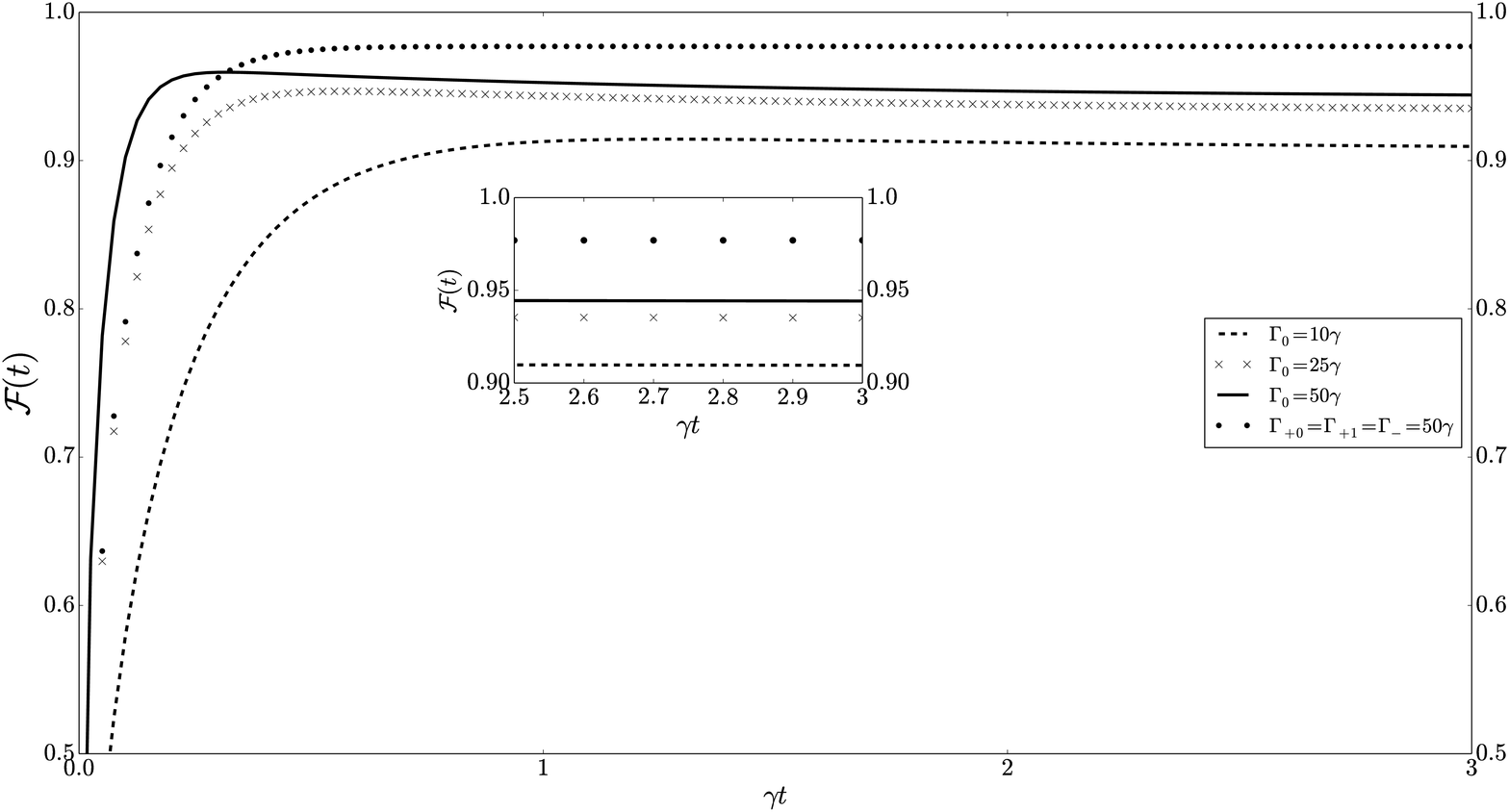}%
\caption{Evolution of the fidelity of
reaching the target state $\left\vert \protect\phi _{+}\right\rangle $,
plotted against the scaled time $\protect\gamma t$, from an initial thermal
state with $\bar{n}=0.05$ in each cavity.}%
\label{fig1}
\end{figure}

To estimate the range of validity of these parameters in a microwave cavity
QED experiment, we start by choosing $\Delta =\Delta _{1}=(1+10^{-2})\times
\Delta _{2}=10\left\vert \Omega _{0}\right\vert $, such that $\left\vert
\Omega _{1}\right\vert =10\times \left\vert \Omega _{2}\right\vert
=\left\vert \Omega _{0}\right\vert $, $\zeta _{0}=10^{-2}\left\vert \Omega
_{0}\right\vert $, $r^{-1}=\tau =10^{2}/\left\vert \Omega _{0}\right\vert $.
Therefore, with typical $\Omega _{0}\sim 5\times 10^{5}$Hz and $\gamma \sim
7.5$Hz for $\bar{n}=0.05$, it follows that $\Gamma _{0}$ has a range up to
the order of $10^{3}\gamma $. Within this regime of parameters we calculate
numerically from Eq. (\ref{7}) the fidelity with which the steady state $%
\left\vert \phi _{+}\right\rangle =(\left\vert 1,0\right\rangle
_{m}+\left\vert 1,0\right\rangle _{m})/\sqrt{2}$ is generated, running in
QuTIP \cite{QuTIP}. In Fig.(\ref{fig1}) we present the evolution of the
fidelity $\mathcal{F}(t)=\sqrt{\func{Tr}\left\vert \phi _{+}\right\rangle
\left\langle \phi _{+}\right\vert \rho (t)}$ for three values of $\Gamma
_{0}=(10,25,50)\gamma $, leading to values around $(0.91,0.93,0.94)$. If we
had chosen $\overline{\omega }_{-}$ instead of $\overline{\omega }_{+}$ to
be almost resonant with the $g\longleftrightarrow i$ transition, we would
have reached the state$\ \left\vert \phi _{-}\right\rangle =(\left\vert
1,0\right\rangle _{m}-\left\vert 1,0\right\rangle _{m})/\sqrt{2}$). We have
also analyzed the effect on the fidelity when the three engineered
Lindbladians act together, with $\Gamma _{+1}=\Gamma _{+0}=\Gamma
_{-}=50\gamma .$ As mentioned before, we achieve a higher fidelity, around $%
0.98.$ The improvement in the preparation of the entangled state is due to
the cooling effect ($\Gamma _{-}$), which enhances the fidelity of the
vacuum state in the mode $\overline{\omega }_{-}.$

\begin{figure}
\includegraphics[width=1.0\columnwidth]{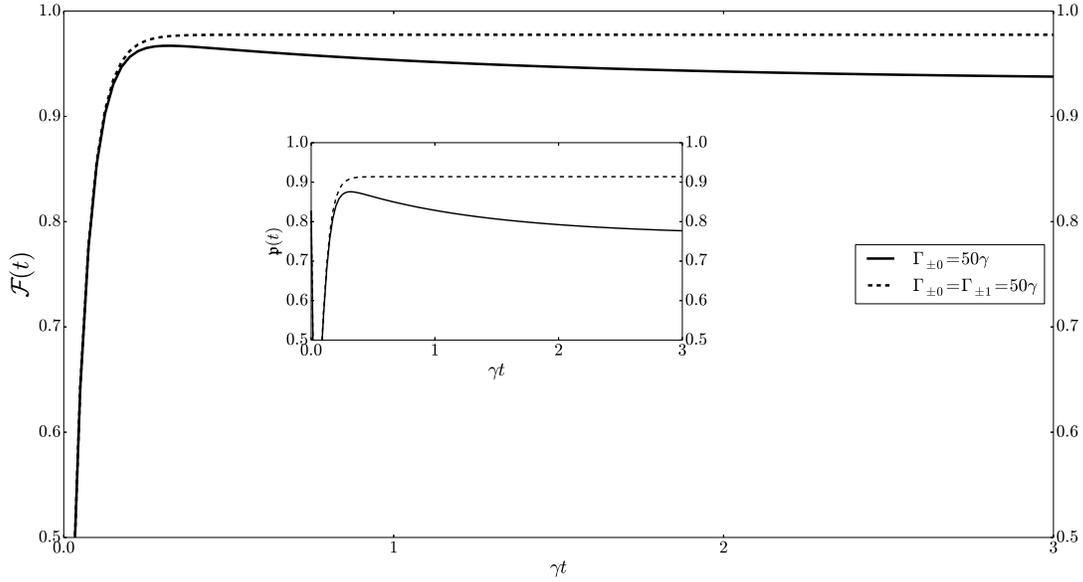}%
\caption{Evolution of the fidelity of
reaching the target state $\left\vert NOON\right\rangle $ against the scaled
time $\protect\gamma t$, from an initial thermal state with $\bar{n}=0.05$
in each cavity. The inset shows the evolution of the purity.}%
\label{fig2}
\end{figure}

In order to investigate the possibility of reaching the state $\left\vert
NOON\right\rangle =$ $(\left\vert 2,0\right\rangle _{m}-\left\vert
0,2\right\rangle _{m})/\sqrt{2}$, we have to consider two atomic beams which
can, for example, each be injected through one of the cavities. We must tune
one of the beams to interact with the normal mode $\overline{\omega }_{+}$
and the other with $\overline{\omega }_{-}$. Following the steps outlined
above to derive master equation (\ref{7}), we reach two selective
Liouvillians acting in space $\left\{ \left\vert 0\right\rangle ,\left\vert
1\right\rangle \right\} $ of the modes $\overline{\omega }_{\pm }$ . In Fig.(%
\ref{fig2}), we present the fidelity $\mathcal{F}(t)=\sqrt{\func{Tr}%
\left\vert NOON\right\rangle \left\langle NOON\right\vert \rho (t)}$ and the
associated purity $\mathfrak{p}(t)=Tr\left[ \rho ^{2}(t)\right] $, achieved
by adopting only the engineered absorption Liouvillian, with $\Gamma _{\pm
0}=50\gamma $, or both the selective absorption and emission Liouvillians,
with $\Gamma _{\pm 0}=\Gamma _{\pm 1}=50\gamma $, leading to fidelities
around $0.93(0.77)$ and $0.98(0.91)$, respectively. In addition to the
increase in fidelity, the use of both selective Liouvillians leads to a
state with a higher degree of purity.

Finally, we investigate the case of degenerate symmetric networks ($\omega
_{m}=\omega $ and $\lambda _{mn}=\lambda )$, where the Hamiltonian (\ref{Hs}%
) can be diagonalized through the canonical transformation $\ A_{1}=\frac{1}{%
\sqrt{N}}\sum_{m}a_{m}$ and $A_{j}=\frac{1}{\sqrt{j(j-1)}}%
\sum_{k=1}^{j-1}a_{k}-(j-1)a_{j}$ ($j=2,3,\ldots ,N)$ and the corresponding
frequencies of the normal modes are $\overline{\omega }_{1}=$ $\omega
+(N+1)\lambda $ and $\overline{\omega }_{j}=$ $\omega -\lambda .\,\ $Here we
are interested in reaching the steady Fock state with a single excitation in
the non-degenerate normal mode $\overline{\omega }_{1}$ and the vacuum state
in all other degenerate modes, which corresponds to a multiqubit W-type
state \cite{WCirac}: $\left\vert 1,0\ldots ,0\right\rangle _{M}=\frac{1}{%
\sqrt{N}}(\left\vert 1,0,\ldots ,0\right\rangle _{m}+\left\vert 0,1,\ldots
,0\right\rangle _{m},\ldots ,\left\vert 0,0,\ldots ,1\right\rangle
_{m})=\left\vert W\right\rangle _{N}.$

\begin{figure}
\includegraphics[width=1.0\columnwidth]{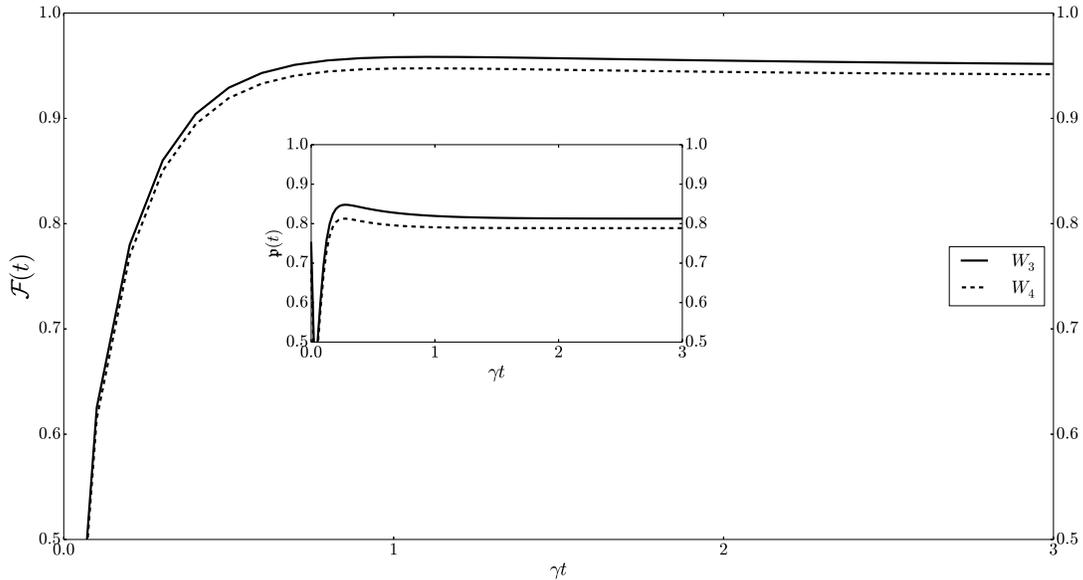}%
\caption{Evolution of the fidelity to
obtain the target states $\left\vert W\right\rangle _{3}$ and\ $\left\vert
W\right\rangle _{4}$ \ against the scaled time $\protect\gamma t$, from a
initial thermal state with $\bar{n}=0.05$ in each cavity. The inset gives
the evolution of the purity.}%
\label{figw}
\end{figure}

We note that the master equation (\ref{eq1}) for the case of degenerate
symmetric networks contains only natural decay rates in the mode $\overline{%
\omega }_{1}$ \cite{NetoJPB}, i.e. $\gamma _{1}=N\gamma $ and $\gamma _{j}=0$%
. Therefore, to reach the target state $\left\vert 1,0\ldots ,0\right\rangle
_{M},$ in addition to the selective Lindbladian for mode $\overline{\omega }%
_{1}$, we need to engineer the cooling Lindbladian in mode $\overline{\omega 
}_{j}$. In a coupled cavity system, we can follow the same steps as those
described above to construct the desired master equation:%
\begin{align}
\frac{d\rho }{dt}& {\small =}\frac{\bar{\gamma}_{10}}{2}\left( 2A_{10}^{\dag
}\rho A_{10}-\rho A_{10}A_{10}^{\dag }-A_{10}A_{10}^{\dag }\rho \right)  
\notag \\
& +(N\gamma /2)(1+\overline{n})\left( 2A_{1}\rho A_{1}^{\dagger }-\rho
A_{1}^{\dagger }A_{1}-A_{1}^{\dagger }A_{1}\rho \right)   \notag \\
& +(N\gamma /2)\overline{n}\left( 2A_{1}^{\dagger }\rho A_{1}-\rho
A_{1}A_{1}^{\dagger }-A_{1}A_{1}^{\dagger }\rho \right) ,  \notag \\
& \dsum_{j}\frac{\tilde{\gamma}_{j}}{2}\left( 2A_{j}\rho A_{j}^{\dagger
}-\rho A_{j}^{\dagger }A_{j}-A_{j}^{\dagger }A_{j}\rho \right) .  \label{sym}
\end{align}%
In Fig.(\ref{figw}) we present the fidelity and purity, computed from Eq.(%
\ref{sym}), of preparation of the target W-type state $\left\vert
W\right\rangle _{N}$, for the cases $N=3,4$, adopting $\Gamma _{10},\Gamma
_{j}=50\gamma $, and starting from a thermal state with $\bar{n}=0.05$. We
find that the fidelities (purities) of the generated entanglements $%
\left\vert W\right\rangle _{3}$ and $\left\vert W\right\rangle _{4}$ are
around $0.95$ $(0.81)$ and $0.94$ $(0.79)$, respectively. In the case of a
degenerate linear network with a single-excitation,\ we can reach a set of
equilibrium multiqubit states given by\ $\left\vert \Psi _{n}\right\rangle =%
\sqrt{(\frac{2}{N+1})}\sum_{k=1}^{N}\sin (\frac{n\pi k}{N+1})a_{k}^{\dagger
}\left\vert 0_{1},...,0_{N}\right\rangle .$

We have thus advanced a theoretical proposal to obtain steady entanglements
in a bosonic dissipative network in the Markovian limit. Our proposal relies
on the engineering of selective JC Hamiltonians, which generate equally
selective Lindblad superoperators that enable us to manipulate the
equilibrium thermal distribution of the normal modes of the network. We also
discuss a possible experimental implementation of our proposal in a system
of coupled cavities where the required engineered Liouvillians are built
from beams of three-level atoms that are made to interact with the network
normal modes.

Addressing some interesting issues to be investigated further, we first
observe that the role played by the network topology in the generation of
the steady genuine multipartite entanglements \cite{ent} was explored only
slightly. Our results indicate that by manipulating the network topology, we
could access a plethora of equilibrium multipartite entanglement states,
covering part or all the network. Finally, it is worth investigating how the
non-Markovianity and the strong interoscillator coupling regime (where the
indirect dissipative channels become effective) affect our dissipative
protocol for preparation of entanglements.

\acknowledgments
The authors acknowledge financial support from PRP/USP within the Research
Support Center Initiative (NAP Q-NANO) and FAPESP, CNPQ and CAPES, Brazilian
agencies.

\end{document}